\documentclass{aipproc}
\layoutstyle{8x11double}

\title
    [The chiral phase transition\ldots]
    {The chiral phase transition and the role of vacuum fluctuations}
\author{Rashid Khan}
{
    address=
    {
        Department of Physics,
        Norwegian University of Science and Technology,
        N-7491 Trondheim,
        Norway
    },
    email={rashid.khan@ntnu.no}
}
\author{Lars T. Kyllingstad}
{
    address=
    {
        Department of Physics,
        Norwegian University of Science and Technology,
        N-7491 Trondheim,
        Norway
    },
    email={lars.kyllingstad@ntnu.no}
}
\keywords{
    Quark-meson effective model,
    optimised perturbation theory,
    chiral symmetry}
\classification{11.10.Wx, 11.25.Db, 11.30.Qc, 11.30.Rd}

\begin{abstract}
    We investigate the chiral phase transition in the quark-meson
    effective model using optimised perturbation theory to one loop.
    Certain terms in the free energy are frequently omitted
    in calculations, on the assumption that their contribution
    is negligible.
    We show that this is not necessarily the case, and that
    the order of the phase transition, as well as
    the critical temperature, depends heavily on which contributions
    are included.
\end{abstract}

\begin{document}
    \maketitle

\section{Introduction}
    The QCD Lagrangian with $N_f$ quark flavours has an
    $SU(N_f)_L \times SU(N_f)_R$ chiral symmetry which is spontaneously
    broken by the formation of a chiral condensate at low temperature
    and baryon density.  The location and nature of the transition
    from the phase with broken chiral symmetry to the chirally
    restored phase is a topic of active research.

    One method of studying the chiral phase transition is by using
    effective models---models which are simpler than QCD, yet share
    QCD's symmetry breaking pattern.  Examples include the linear
    sigma model, NJL models, and the quark-meson (QM) effective model
    that we will use in the present work.

    Effective models are by their very nature simplifications of a
    more complicated theory.  Even so, one often makes further
    simplifications within the models themselves.  One may for
    instance decide to simply omit certain parts of the model
    that don't seem immediately important for the problem at hand.
    As an example, in the QM model one often neglects the vacuum
    contribution to the free energy from the quarks, since the
    chiral symmetry breaking takes place in the meson sector.
    However, one should perhaps not make such assumptions light-handedly.
    It was recently shown that the choice of whether or
    not to include the fermion vacuum contribution actually
    changes the order of the phase transition \cite{Skokov:2010sf}.

    In the following, we will look at various approximations
    to the free energy of the QM model, and discuss their impact
    on the phase diagram.

    We will calculate the free energy using \emph{optimised
    perturbation theory} (OPT) \cite{Chiku:1998kd}, which is
    a framework for reorganising the perturbative series to obtain
    improved convergence.
    The present discussion will be limited to the one-loop case,
    as the two-loop calculation is still a work in progress.

\section{The model}
    We consider the quark-meson effective model with $N$ scalar fields
    ($\phi_1, \phi_2, \ldots, \phi_N$) and $N_f$ massless quark flavours
    ($\psi$).  The Euclidean Lagrangian can be written as
    \begin{equation}
        \mathcal L =
            \mathcal L_\mathrm{m}
          + \mathcal L_\mathrm{q}
          + \mathcal L_\mathrm{i},
    \end{equation}
    where the three terms
    describe the mesons, the quarks, and the interactions between them,
    respectively.  They are defined by:
    \begin{eqnarray}
        \mathcal L_\mathrm{m} &=&
            \frac{1}{2} (\partial_\mu \phi_i) (\partial_\mu \phi_i)
          + \frac{1}{2} m_0^2 \phi_i \phi_i
          + \frac{\lambda}{4!} (\phi_i \phi_i)^2,
          \label{eq:lagr_lsm}
        \\
        \mathcal L_\mathrm{q} &=&
            \bar\psi (\gamma_\mu \partial_\mu - \mu \gamma_4) \psi,
        \\
        \mathcal L_\mathrm{i} &=&
            g \bar\psi \Gamma_i \psi \phi_i.
    \end{eqnarray}
    Here, we have defined $\Gamma \equiv (1, -i \gamma_5 \tau)$,
    where $\tau = (\tau_1, \tau_2, \tau_3)$ are the Pauli
    matrices\footnote{
        This only makes sense when $N=4$, which is the value we will
        use in numerical calculations.
    }. $\mu$ is the quark chemical potential, which we assume to be equal
    for all quark flavours, and which is therefore related to the baryon
    chemical potential by $\mu = \mu_B/3$.

    When $N = 4$ and $N_f = 2$, this Lagrangian has a
    $SU(2)_L \times SU(2)_R$ symmetry which, assuming $m_0^2 < 0$,
    is spontaneously broken down to $SU(2)_V$ at low energy\footnote{
        In the scalar sector, the symmetry manifests itself as
        $O(N)$, spontaneously broken down to $O(N-1)$.
    }.
    In the broken phase, $\phi$ acquires a nonzero expectation value $v$,
    called the \emph{chiral condensate}, and we therefore write
    \begin{equation}
        \phi = (\sigma + v, \pi_1, \pi_2, \ldots, \pi_{N-1}).
    \end{equation}

    In optimised perturbation theory, we add and subtract quadratic
    terms in the Lagrangian:
    \begin{equation}
        \mathcal L \to
            \mathcal L
          + \frac{1}{2} \chi (\sigma^2 + \pi_i \pi_i)
          - \frac{1}{2} \chi (\sigma^2 + \pi_i \pi_i).
    \end{equation}
    The terms with positive sign are treated as mass terms, while
    the terms with negative sign are treated as two-particle
    interactions that give rise to a set of new Feynman rules.
    In this way we obtain a systematic resummation of selected diagrams
    from \emph{all} orders of the na\"ive perturbation expansion.

\section{Free energy}
    At tree level, we find that the only
    contribution to the free energy comes from the meson sector:
    \begin{equation}
        \mathcal F_0 =
            \frac{1}{2} m^2 v^2
          + \frac{\lambda}{4!} v^4.
    \end{equation}

    At one loop, after calculating the relevant
    Feynman diagrams in dimensional regularisation and performing
    renormalisation according to the $\overline{\rm MS}$ scheme,
    we identify the following contributions to the free energy:
    \begin{equation}
        \mathcal F_1 =
            \mathcal F_\mathrm{OPT}
          + \mathcal F_b^\mathrm{vac}
          + \mathcal F_b^T
          + \mathcal F_f^\mathrm{vac}
          + \mathcal F_f^T.
        \label{eq:free_1}
    \end{equation}
    The first one, $\mathcal F_\mathrm{OPT}$, arises from the OPT
    interaction term, and is defined as
    \begin{equation}
        \mathcal F_\mathrm{OPT} = - \frac{1}{2} \chi v^2.
    \end{equation}
    Next, we have the vacuum (i.e. $T=0$) contribution from the
    scalar fields:
    \begin{eqnarray}
        \mathcal F_b^\mathrm{vac} &=&
          - \frac{\delta}{4 (4\pi)^2}
            \left[
                \frac{3}{2}
              - 2 \log \frac{m_\sigma}{\Lambda}
            \right]
            m_\sigma^4
            \nonumber \\ &&
          - \frac{\delta (N-1)}{4 (4\pi)^2}
            \left[
                \frac{3}{2}
              - 2 \log \frac{m_\pi}{\Lambda}
            \right]
            m_\pi^4.
    \end{eqnarray}
    Here, $\Lambda$ is the $\overline{\rm MS}$ renormalisation scale.
    
    The thermal (i.e. temperature-dependent) contribution from the
    meson sector is given by
    \begin{eqnarray}
        \mathcal F_b^T &=&
          - \frac{\delta 8}{3 (4\pi)^2}
            \int_0^\infty dp
            \frac{p^4}{\omega_\sigma(p)}
            n_B(\omega_\sigma(p))
            \nonumber \\ &&
          - \frac{\delta 8 (N-1)}{3 (4\pi)^2}
            \int_0^\infty dp
            \frac{p^4}{\omega_\pi(p)}
            n_B(\omega_\pi(p)),
    \end{eqnarray}
    where $\omega_a(p) = \sqrt{p^2 + m_a^2}$ and $n_B$ is the
    Bose-Einstein distribution function.

    Finally, we have the vacuum and thermal contributions from the
    quark sector, which are given by, respectively,
    \begin{eqnarray}
        \mathcal F_f^\mathrm{vac} &=&
            \frac{\delta N_f}{(4\pi)^2}
            \left[
                \frac{3}{2}
              - 2 \log \frac{m_q}{\Lambda}
            \right]
            m_q^4,
        \\
        \mathcal F_f^T &=&
          - \frac{\delta 16 N_f}{3 (4\pi)^2}
            \int_0^\infty dp
            \frac{p^4}{\omega_q(p)}
            \nonumber \\ && \times
            \left[
                n_F(\omega_q(p) + \mu)
              + n_F(\omega_q(p) - \mu)
            \right],
    \end{eqnarray}
    where $n_F$ is the Fermi-Dirac distribution function and $m_q$
    is a dynamically generated quark mass, $m_q = g v$.

\section{Results}
    The optimisation parameter $\chi$ is in principle completely
    arbitrary, and if we could carry out calculations to \emph{all}
    orders, the result would be independent of $\chi$.  As it stands,
    we are forced to truncate the loop expansion at a certain order,
    and we must therefore find a prescription for $\chi$.

    To that end, we consider the gap equation for the chiral condensate
    $v$, obtained by differentiating the free energy:
    \begin{equation}
        0
        =
        \frac{d}{dv} (\mathcal F_0 + \mathcal F_1)
        =
        v (m_\pi^2 + \Pi_1)
        \label{eq:gapeq}
    \end{equation}
    Here, $\Pi_1$ is the one-loop self-energy
$        \Pi_1 = \frac{1}{v} \frac{d \mathcal F_1}{dv}$.

    We now apply a \emph{fastest apparent convergence} (FAC) criterion
    to $\chi$:  We choose $\chi$ so that the perturbative correction
    $\Pi_1$ is as small as possible.  In fact, we can require
    that it vanishes completely, $\Pi_1 = 0$.
    Then it is only a matter of solving Eq.\ \eqref{eq:gapeq},
    now simplified to $v m_\pi^2 = 0$, for $v$.

    We will now look at some common approximations to the free energy
    $\mathcal F$, each obtained by neglecting some of the terms in
    Eq.\ \eqref{eq:free_1}, and we will demonstrate the effect of each
    approximation on the structure of the phase diagram.
    Lattice calculations on two-flavour QCD have placed the phase
    transition at $T_c \sim 150$ MeV for $\mu_B = 0$, and shown
    that it is of second order.  This can, to some extent,
    be used as a measure of the success of the various
    approximations---at least qualitatively.

\begin{figure}
        \includegraphics[width=7cm]{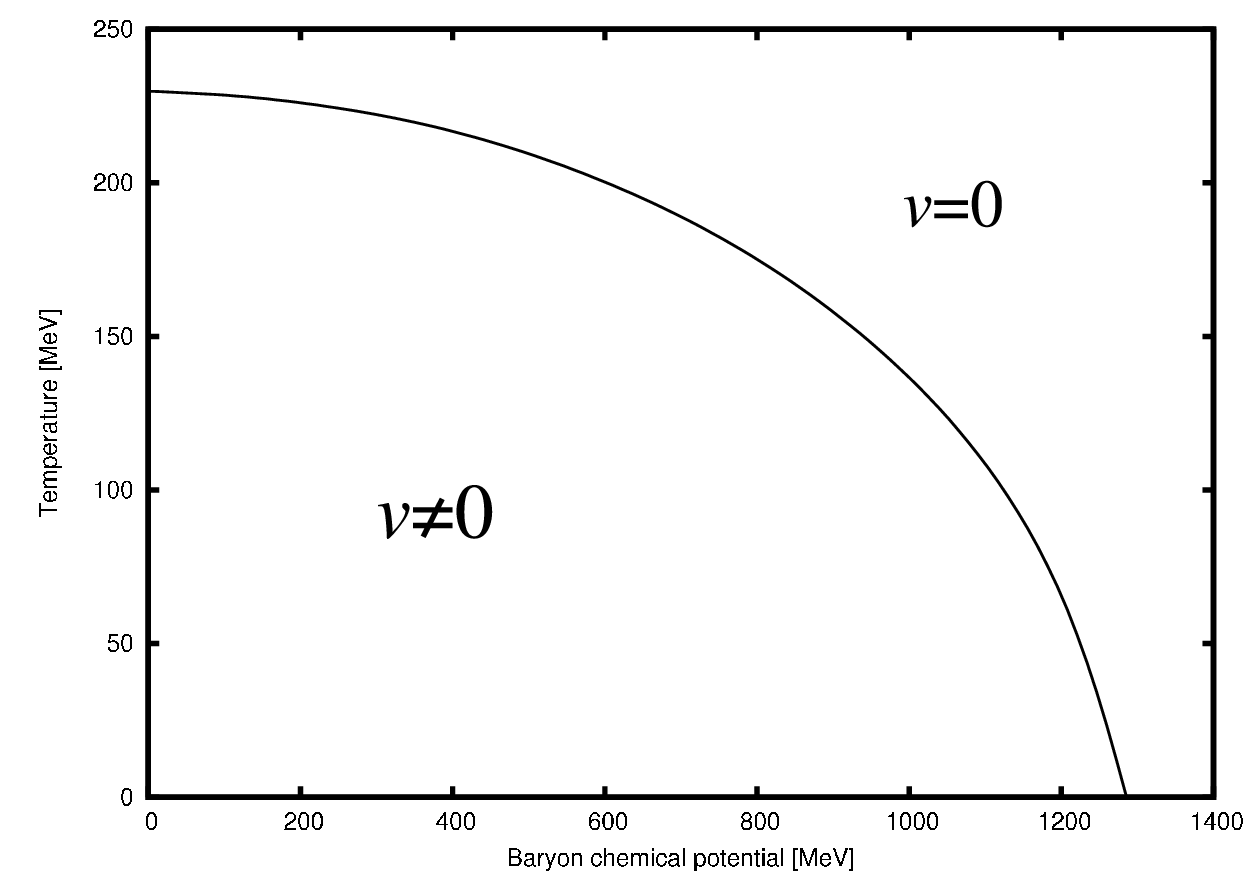}
        \caption{$\mathcal F = \mathcal F_0 + \mathcal F_\mathrm{OPT}
            + \mathcal F_f^T$.}
        \label{fig:no-vacuum_no-bosons}
    \end{figure}

    Starting with the simplest case, we may say that we are only
    interested in thermal effects, and thus ignore the vacuum contributions
    $\mathcal F_b^\mathrm{vac}$ and $\mathcal F_f^\mathrm{vac}$.  Furthermore,
    it is common to use a mean-field approximation to the meson sector,
    in which we ignore the thermal fluctuations, $\mathcal F_b^T$, as well.
    We show the phase diagram one obtains with this approximation in
    Fig.\ \ref{fig:no-vacuum_no-bosons}.  The solid line indicates a first-order
    phase transition, and the critical temperature for $\mu_B$ is
    rather high, $T_c = 229$ MeV.

\begin{figure}
        \includegraphics[width=7cm]{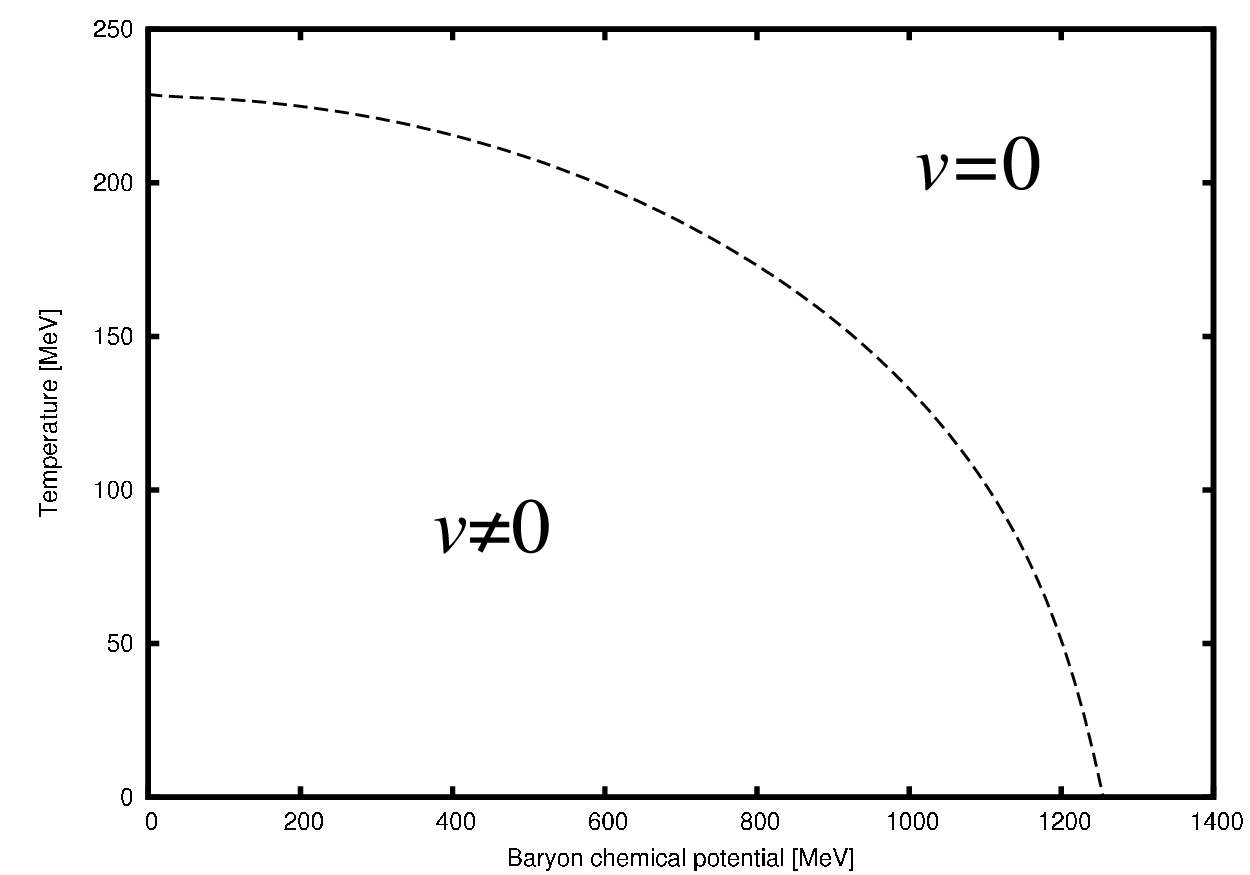}
        \caption{$\mathcal F = \mathcal F_0 + \mathcal F_\mathrm{OPT}
            + \mathcal F_f^\mathrm{vac} + \mathcal F_f^T$.}
        \label{fig:no-bosons}
    \end{figure}

    For $\mu_B = 0$, it was recently shown by Skokov {\it et al} that if one,
    in addition to the above, includes the  fermion vacuum term,
    $\mathcal F_f^\mathrm{vac}$, the order of the phase
    transition actually changes---it becomes a second-order phase transition
    \cite{Skokov:2010sf}.
    We have found this to be true for nonzero $\mu_B$ as well, as
    shown in the phase diagram in Fig.\ \ref{fig:no-bosons}.  The dashed
    line indicates a second-order phase transition.
    The critical temperature doesn't change much compared to the above.

\begin{figure}
        \includegraphics[width=7cm]{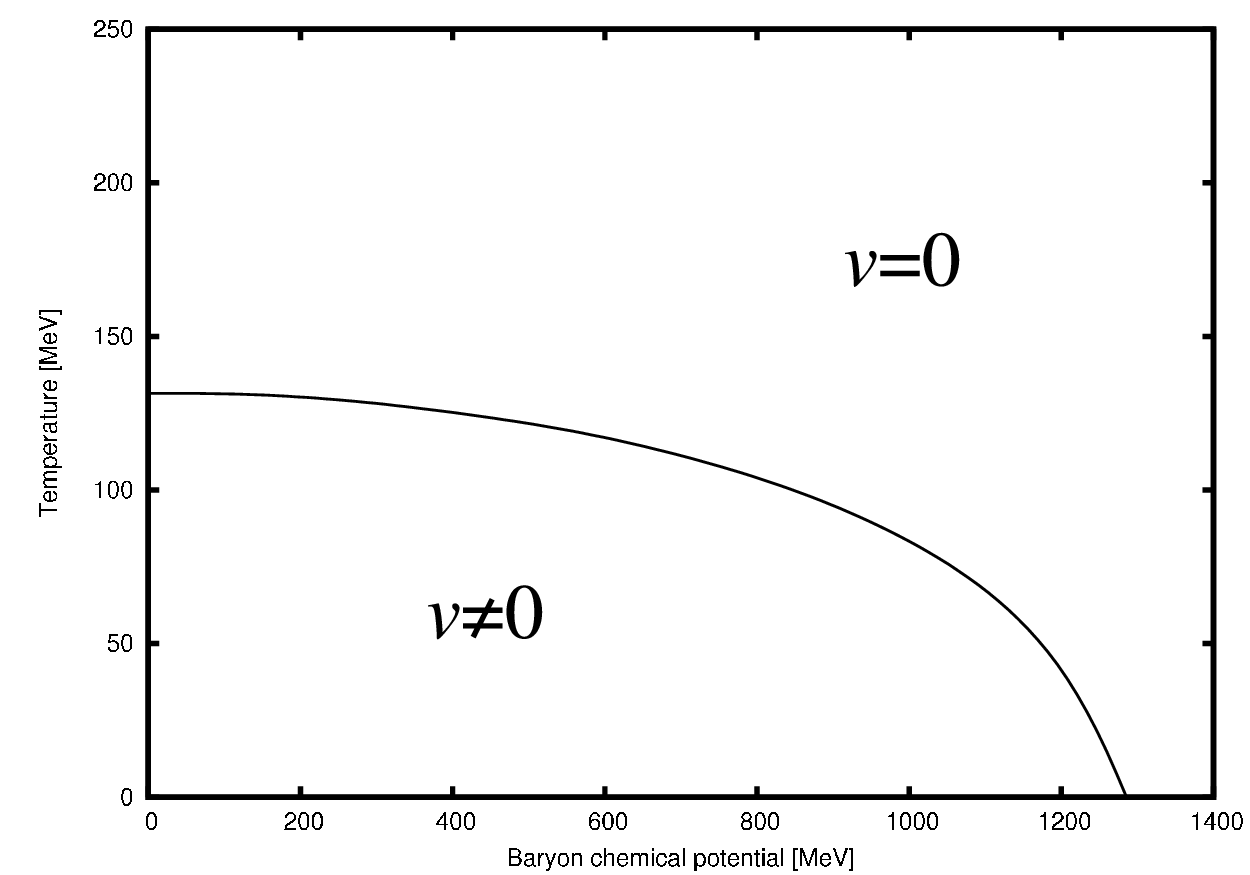}
        \caption{$\mathcal F = \mathcal F_0 + \mathcal F_\mathrm{OPT}
            + \mathcal F_b^T + \mathcal F_f^T$.}
        \label{fig:no-vacuum}
    \end{figure}

    Next, we again disregard the fermion vacuum fluctuations, and instead
    we include the thermal contibution from the meson sector.  This will
    allow us to see the effect of taking the mean-field approximation
    in the first place.  The phase diagram is shown in
    Fig.\ \ref{fig:no-vacuum}, and it agrees very well with the results
    of Bowman and Kapusta \cite{Bowman:2008kc}. we see that the
    phase transition is first order, and that the critical temperature
    is now significantly lower.  Now, $T_c = 132$ MeV, which is much
    closer to the QCD critical temperature.

\begin{figure}
        \includegraphics[width=7cm]{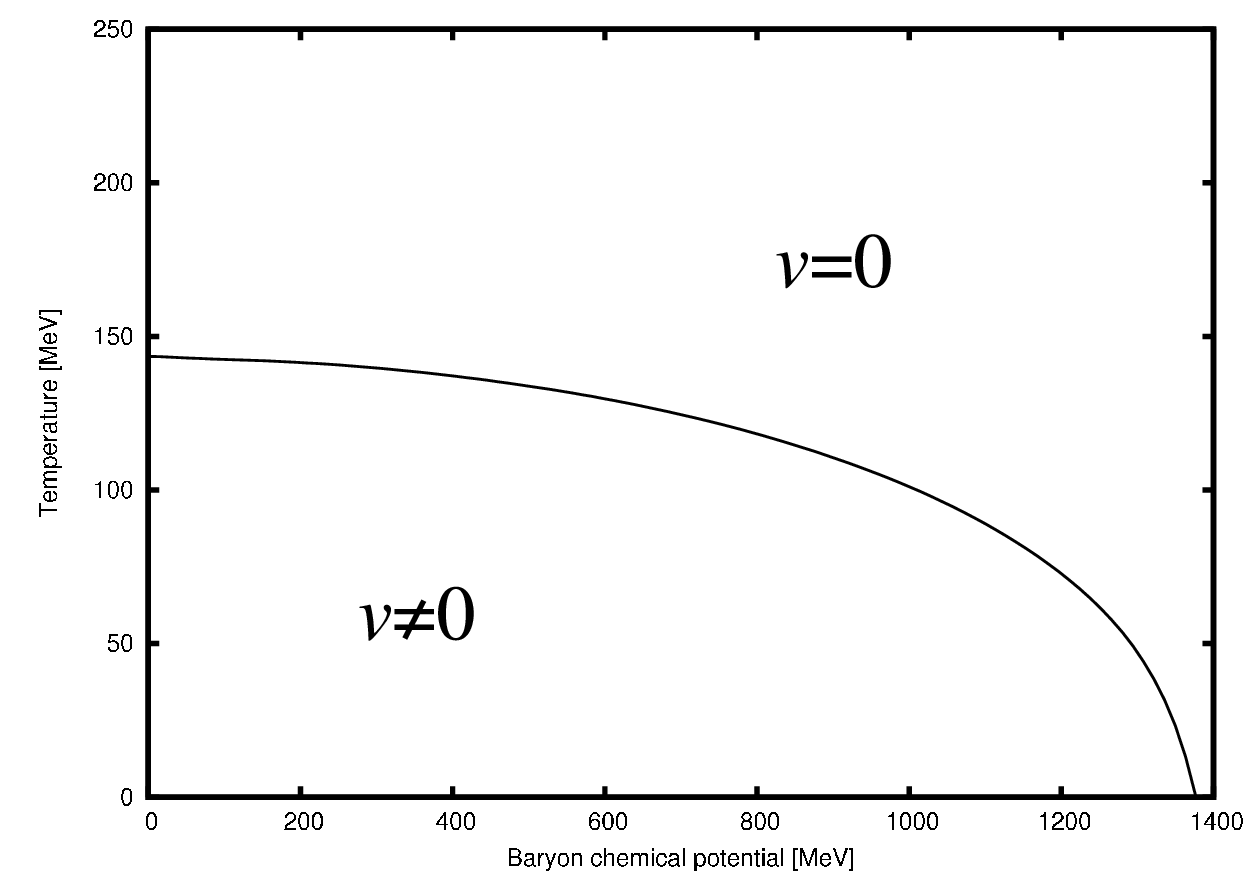}
        \caption{$\mathcal F = \mathcal F_0 + \mathcal F_\mathrm{OPT}
            + \mathcal F_b^\mathrm{vac} + \mathcal F_b^T
            + \mathcal F_f^\mathrm{vac} + \mathcal F_f^T$.}
        \label{fig:full}
    \end{figure}

    Finally, in Fig.\ \ref{fig:full}, we show the phase diagram when
    all contributions to the free energy are taken into account.  The
    phase transition is first order for all $\mu_B$, and the critical
    temperature is $T_c = 142$ MeV.

\section{Summary and outlook}
    We started with a simple approximation, where we only included
    the tree-level contribution from the bosons and the thermal
    contribution from the fermions, and got a first-order phase
    transition with a rather high critical temperature.

    We then saw that we could add the vacuum contribution from the
    fermions, which resulted in the phase transition becoming a
    second-order phase transition, or we could add the thermal
    contribution of the bosons, which pushed the critical temperature
    down into the vincinity of the QCD critical temperature.

    It would be tempting to guess that by including all contributions,
    we would get a second-order phase transition with a low critical
    temperature, but as we have seen, that is not the case.
    When the vacuum fluctuations of both fermions and bosons are
    included, they have a significantly weaker impact on the
    phase diagram.  Part of the reason for this is the fact that the
    vacuum energy of the mesons has opposite sign from the vacuum
    energy of the quarks, so to a certain degree they simply cancel
    out.

    The next step is obviously to include the next loop order.
    At two-loop order 
    we expect to capture more aspects of the QCD phase diagram,
    in particular we expect the phase transition to become second
    order at high $T$ and low $\mu_B$.
    This is work currently in progress.

\bibliographystyle{aipproc}
\bibliography{proceedings}
\end{document}